\documentclass[pra,aps,twocolumn]{revtex4-1}
\usepackage{newlfont}
\usepackage{amssymb}
\usepackage{amsfonts}
\usepackage{amsmath}
\usepackage{wasysym}
\usepackage{graphicx}
\usepackage{epsfig}
\usepackage{amsthm}
\usepackage[usenames]{color}           
\usepackage{soul}            
\usepackage[colorlinks=true, citecolor=blue, urlcolor=blue ]{hyperref}

\newcommand{\be}{\begin{equation}}
\newcommand{\ee}{\end{equation}}
\newcommand{\bqs}{\begin{equation*}}
\newcommand{\eqs}{\end{equation*}}
\newcommand{\ba}{\begin{array}}
\newcommand{\ea}{\end{array}}
\newcommand{\bas}{\begin{array*}}
\newcommand{\eas}{\end{array*}}
\newcommand{\bqa}{\begin{eqnarray}}
\newcommand{\eqa}{\end{eqnarray}}
\newcommand{\bqas}{\begin{eqnarray*}}
\newcommand{\eqas}{\end{eqnarray*}}

\begin{document}



\title{Benford's law gives better scale exponents in phase transitions of quantum XY models}

\author{Ameya Deepak Rane, Utkarsh Mishra, Anindya Biswas, Aditi Sen(De), and Ujjwal Sen}

\affiliation{Harish-Chandra Research Institute, Chhatnag Road, Jhunsi, Allahabad 211 019, India}

\begin{abstract}
Benford's law is an empirical law predicting the distribution of the first significant digits of numbers
obtained from natural phenomena and mathematical tables.
It has been found to be applicable for numbers coming from 
a plethora of sources,
 varying from seismographic, biological, financial, to astronomical. 
We apply this law to analyze the data obtained from physical many-body systems described by the
one-dimensional anisotropic quantum XY models in a 
transverse magnetic field. We detect the zero-temperature quantum phase transition
and find that our method gives better finite-size scaling exponents for the critical
point than many other known scaling exponents using measurable quantities like magnetization, entanglement, and quantum discord. 
We extend our analysis to the same system but at finite temperature and 
find that it also detects the finite temperature phase transition in the model. 
Moreover, we compare the Benford distribution analysis with the same obtained from
the uniform and Poisson 
distributions.
The analysis is furthermore important in that the high-precision detection of the 
cooperative physical phenomena
is possible even from low-precision experimental data.

\end{abstract}

\maketitle

\section{Introduction}
Benford's law is an empirical law, based on the observation that the first significant digits, 
in a given set of data 
are not random. 
Instead, the chance that the first significant digit from the set of data happens to be
``1'' is almost $30$\%, ``2'' almost 17\%, and so on. 
The frequency of the digit ``9'' is the lowest among all the digits from $1$ to \(9\), and it is around $4.5$\%.
The law was first observed by the astronomer Simon Newcomb in 1881~\cite{newcomb}, who also 
proposed the probability function 
to the distribution. It is this distribution which is today known as the ``Benford's Law'' or 
``The Law of First Digit'' or 
``The Leading Digit Phenomenon''. This law was rediscovered by Frank Benford in 1938~\cite{benford}. 
Benford's law predicts the frequency distribution of the first significant digit \(D\) as
\begin{equation}
  P_{D}=\log_{10}(1+\frac{1}{D}).
\end{equation}
The law has since been verified for a wide spectrum of situations in natural science, and  
while studying mathematical series~\cite{record}.  
Not all sets of numbers obey this law, e.g. random numbers obtained from a computer 
code will not obey such a law, and 
the frequency of any number from $1$ to \(9\) as 
a leading digit will be $1/9$. But there are examples ranging from data obtained in earthquakes, and the 
brightness of gamma rays that reaches earth, to
the rotation rates of spinning stars, 
where Benford's law is respected~\cite{record}. Violations of the law have been used in 
detecting cases of tax fraud, election
 fraud, digital image manipulation, faint earthquakes, and phase transitions~\cite{stephenBattersby,au_benford}. 
The scale invariance of Benford law also makes it independent of the measuring device used to get the source data~\cite{scale_inv}.

A quantum phase transition (QPT) is a change in the phase of a system at zero temperature, driven by some parameter of the system, 
like external magnetic field or the coupling strength between particles. Unlike the thermal phase 
transitions~\cite{classical-Sachdev} where the transition arises due to thermal fluctuations in the system, 
 quantum phase transitions are driven solely by quantum fluctuations~\cite{Ssachdev}.
Understanding  quantum phases in many-body 
quantum systems is important for a number of reasons, including discerning quantumness in systems of several 
particles and realizing quantum computers.
The one-dimensional transverse XY model is an integrable model where a quantum phase 
transition occurs at zero temperature when the external transverse field is varied~\cite{LSM,BMD,BM}. This model 
has been studied, e.g. in 
crystals of CoNb$_{2}$O$_{6}$~\cite{expt2}. Spectacular advances in cold gas experimental techniques in recent years have led to the 
possibility of experimental detection of such transitions in optical lattice systems and ion traps~\cite{Rislam}.

In this paper, we consider the anisotropic quantum XY models, in one dimension, for finite- and infinite-size systems. 
 The system undergoes a quantum phase transition at
$\lambda \equiv h/J=\lambda_c \equiv 1$, where $h$ is the strength of the external magnetic field while $J$ is the coupling 
strength between neighboring spins. Here we use the Benford's law to detect the transition and 
investigate its scaling properties.
The main advantage of this approach is its amenability to experiments, owing to the fact that one requires to investigate only the 
first significant digit of an observable, which can be easily measured in experiments. 
We study the transverse magnetization as a function of the external magnetic field.
The difference between the frequency distributions of the first significant digits of the transverse magnetization 
and the expected frequency distribution from Benford's law, is quantified using a parameter called the Benford violation parameter~(BVP).
The maxima of the derivative of the BVP with respect to the magnetic field, indicates the point of the quantum phase transition 
($\lambda_c^N$), in
finite size systems. Finite size scaling analysis shows that the phase transition point for finite systems ($\lambda_c^N$)
approaches the critical point for infinite systems ($\lambda_c$) as
$N^{-\alpha}$, where $\alpha$ can be as high as -2.45, and $N$ is the number of spins. 
We further extend this analysis to the infinite-size quantum XY model, 
but at non-zero temperature. The BVP shows two crossovers in the $\lambda-\widetilde{T}$ plane,
as the magnetic field is varied across the critical point, where $\widetilde{T}$ is a scaled temperature. 
The finite temperature quantum critical region, for unit anisotropy, 
is contained within the lines
$\widetilde{T}=-0.546(\lambda-\lambda_{c})$, for
$\lambda<\lambda_c$ and $\widetilde{T}=0.567(\lambda-\lambda_{c})$, for $\lambda>\lambda_c$.
Besides, we find that the Benford distribution is not the only distribution which can be used
to detect the quantum phase transition. We show that other discrete distributions like the 
uniform and the Poisson (for a number of values of its parameter) can signal the QPT. 
We find the scaling exponents of the quantum critical point by
using the uniform and the Poisson 
distributions. We also analyze the effect of changing the distance measure between the observed and 
predicted probability distributions
on the scaling exponents. The distance measures used here are mean deviation, standard deviation, and the 
Bhattacharya metric~\cite{Bhattacharya}. We find that the analyses using the Benford distribution, in general, 
provide higher values of the scaling exponents.

The paper is organized as follows. In Sec.~\ref{sec:model_def}, we briefly discuss the one-dimensional quantum XY models
for both finite and infinite systems.
 Sec.~\ref{sec:bvp} contains a discussion on the methodology and tools used in the analysis of the data.
The detection of phase transition by using the BVP of transverse magnetization and the scaling of quantum 
critical point for finite size systems is discussed in Sec.~\ref{sec:finite_system}. Further analysis of the phase space diagram in the 
finite temperature case for infinite systems is taken up 
in Sec.~\ref{sec:finite_temp}. In Sec.~\ref{sec:compare}, we show that the uniform 
and the Poisson distributions can also be used to detect a QPT. We compare the results
with the ones obtained using the Benford distribution. Finally, we conclude in Sec. \ref{sec:conclusion}.

\section{Description of the model}
\label{sec:model_def}
The Hamiltonian of the one-dimensional anisotropic quantum XY model is given by
\begin{equation}
 H = \dfrac{J}{2} \displaystyle \sum_{i=1}^N \left[(1+\gamma)\sigma_{i}^{x}\sigma_{i+1}^{x}
 +(1-\gamma)\sigma_{i}^{y}\sigma_{i+1}^{y}\right] +h \sum_{i=1}^N \sigma_{i}^{z}
 \label{eq:ising_H}
\end{equation}
where $J$ is the coupling constant, 
$h$ is the strength of the transverse magnetic field, $\gamma~(\ne0)$ is the anisotropy parameter,
and $\sigma$'s are the Pauli matrices in a system of $N$ quantum spin-1/2 particles.
We assume periodic boundary condition.
The system undergoes 
a quantum phase transition from long range antiferromagnetic to paramagnetic phase at $h/J=1$~\cite{LSM, BMD,BM}.  
The model is diagonalizable by applying successive Jordan-Wigner, Fourier, and Bogoliubov transformations~\cite{LSM,BMD,BM}.
The average transverse magnetization can be calculated for any number of spins at any temperature. 
The anisotropic quantum XY models for $\gamma \ne 0$ forms the ``Ising universality class''. 
For $\gamma=1$, the Hamiltonian described by Eq.~(\ref{eq:ising_H}) is known as the Ising Hamiltonian.
For finite spin systems,
the transverse magnetization reads
 \begin{equation}
	M_{z}(\lambda,\widetilde{\beta},N)=-\frac{2}{N}\sum_{p=1}^{N/2}\frac{\tanh(\widetilde{\beta}\Lambda(\lambda)/2)
	(\cos(\phi_{p})-\lambda)}{\Lambda(\lambda)},
	\label{eq:mz_finite}
\end{equation}
where  $\beta=\frac{1}{kT}$, $k$ is the Boltzmann constant, $T$ is the absolute temperature, 
$\widetilde{\beta}=\beta J$, $\phi_{p}=\frac{2\pi p}{N}$,
and $\Lambda (x)=\left\{\gamma^2\sin^2(\phi_p)+[x-\cos(\phi_p)]^{2}\right\}^{1/2}$,
while for infinite systems, the transverse magnetization is given by
\begin{equation}
\label{eq:fintemp}
	M_{z}(\lambda,\widetilde{\beta})=-\frac{1}{\pi}\int\limits_{0}^{\pi}d\phi\frac{\tanh(\widetilde{\beta}\Lambda
	(\lambda)/2)(\cos(\phi)-\lambda)}{\Lambda(\lambda)}
\end{equation}
where $\Lambda (x)=\left\{\gamma^2\sin^{2}(\phi)+[x-\cos(\phi)]^{2}\right\}^{1/2}$.
The two site correlation functions can also be calculated analytically for this model for both finite and
infinite lattice sizes at any temperature. The nearest neighbor diagonal correlations can be expressed in 
terms of a correlator, $G(R,\lambda)$, through
 \begin{equation}
   C_{xx}(\lambda)=G(-1,\lambda), C_{yy}(\lambda)=G(1,\lambda),
 \label{eq:cxx}
 \end {equation}
 and
 \begin{equation}
 	C_{zz}(\lambda)=[M_{z}(\lambda)]^{2}-G(-1,\lambda)G(1,\lambda),	
 \end {equation}
 where $G(R,\lambda) $, for infinite lattice size and zero temperature, is given by  
 \begin{equation}
 	G(R,\lambda)=\frac{1}{\pi}\int\limits_{0}^{\pi}d\phi\frac{(\gamma\sin(\phi R)\sin(\phi)-\cos(\phi)(\cos(\phi)-\lambda))}{\Lambda(\lambda)}.
 	\label{eq:GR}
 \end{equation}

\section{Benford Violation Parameter: the methodology}
\label{sec:bvp}
We now discuss the methodology employed to analyze the data obtained for 
a given observable by 
using the Benford violation parameter.
The idea of the Benford violation parameter 
is to characterize an observable in terms of the frequencies of the first significant digits. 
To do this, we  compare it with the expected Benford frequency, 
and quantify it with a number, which we call  the Benford violation parameter.

\subsection{Computing the BVP}
For an observable $Q(x)$, defined in a range $[a,b]$ of $x$, we sample $N$ points in the range. 
Let us denote the minimum and maximum values of the observable in a subinterval $[a',b']$ comprising of $n$ points, 
of the total interval $[a,b]$, by $Q_{\min}$ and $Q_{\max}$ 
respectively. Using these two values, we create a set of data
for the observable, such that all the values in $[a',b']$ lie in the range $[0,1]$~\cite{au_benford}. 
The new value of $Q$, ``Benford $Q$'', is denoted by $Q^{B}$, and is given by
\begin{equation}
\label{eq:nor_benf}
 Q^{B}=\frac{Q-Q_{\min}}{Q_{\max}-Q_{\min}}.
\end{equation}
The frequency of the first significant digit to be D, obtained from these new values, is called the ``observed'' 
frequency, $O_{D}$, for D=1,...,9.
Note that the procedure given in Eq.~(\ref{eq:nor_benf}) is important to
obtain a nontrivial frequency distribution
of digits from $1$ to $9$.  
The rescaling is necessary, since a distribution varying between 1 and 2, for example, will never have 
the first significant digit larger than 2.
The next step is to compare $O_{D}$ with 
the expected frequency distribution given by Benford's law, denoted by $E_{D}=n\log_{10}(1+\frac{1}{D})$.
So, for any observable $Q$, we denote the violation parameter by $\delta(Q)$, and define it as
\begin{equation}
	\delta(Q)=\sum_{D=1}^{9}\left|\frac{O_{D}-E_{D}}{E_{D}}\right|.
	\label{eq:bvp}
\end{equation}
This number gives us the variation of the observable with respect to the Benford frequency and we
assign this number to the mid-point of the range $[a',b']$. The subinterval $[a',b']$ determines the error of the
variable $x$. Therefore, it should be small compared to the range of the total interval $[a,b]$. The number of data
points $n$ in this subinterval should be large enough to ensure convergence of the BVP.
Note that, lower the value of $\delta(Q)$, better is the distribution's conformity with the Benford's law.

\section{Benford scaling of quantum phase transition in the transverse XY model}
\label{sec:finite_system}
In this and the succeeding two sections, we present our main findings regarding the BVP and quantum phase transition in the 
quantum XY model. 
The current section deals with the zero temperature behavior, while the succeeding one is for the finite-temperature regime.
In the zero temperature case, we first show
 that the BVP of the transverse magnetization and correlations  are able to detect the QPT in this model.
 We subsequently perform a finite-size scaling analysis on the BVP data of the system, which indicates the possibility of 
detecting the QPT in finite-sized systems, potentially realized in cold gas experiments. 
 We show that the scaling exponent obtained by using the BVP 
of $M_z$ is much higher than the ones obtained by using $M_z$ itself and several other physical quantities.
The analysis points to the following interesting possibility. By using the BVP as an order parameter, a 
QPT can be detected with high-precision even in finite-sized systems, and even in cases where the observables 
can be measured in the experiment with a low precision. 

\subsection{Detection of QPT}
\label{anisoXY}
\begin{figure}[]
\includegraphics[width=0.48\textwidth]{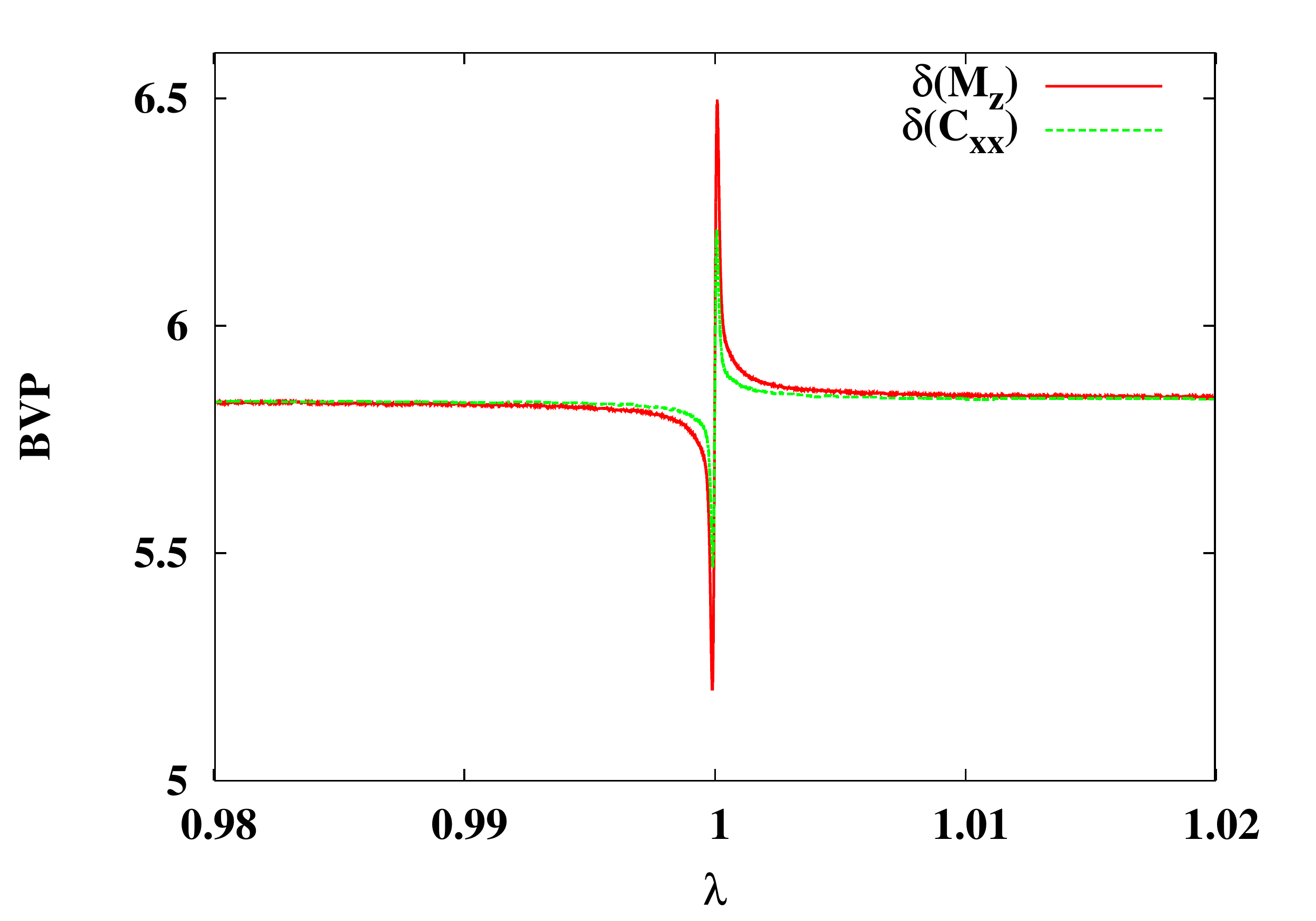} {\centering}
\caption{(Color online.) The BVP for  $M_z$ and $C_{xx}$ for the infinite  quantum Ising model.
There is an abrupt change in the BVP at the QPT point. The axes are dimensionless. The features remain similar for any 
\(\gamma \ne 0\).}
\label{fig:mzc}
\end{figure}

We now compute the BVP for $M_z$ and $C_{xx}$ for the infinite quantum XY model. 
The basic idea involved in Benford analysis for any 
observable was discussed in the preceding section. 
To obtain the Benford 
magnetization, $M_{z}^{B}$, for a given value of the driving parameter
$\lambda$, we choose a 
small interval around $\lambda$ of width $\epsilon$, 
$(\lambda-\epsilon/2,\lambda+\epsilon/2)$,
where $\epsilon$ is a small number. In this small interval, we choose $n$ values of
the system parameter, $\lambda$. Corresponding to those $n$ values of the field, we 
get $n$ values for the transverse magnetization of the system from 
Eq.~(\ref{eq:fintemp}). From this set of data for
magnetization, the normalized Benford transverse magnetization 
is obtained by using Eq.~(\ref{eq:nor_benf}). 
To get the violation parameter for \(M_z\),
we then find the corresponding $O_{D}$'s and subsequently use Eq.~(\ref{eq:bvp}).
The same analysis is performed in the
entire range of $\lambda$. The entire procedure is repeated for the calculation of 
BVP of $C_{xx}$. We find that away 
from the quantum critical point ($\lambda_{c}=1$), 
the BVP is almost constant and changes little as we change $\lambda$ (see Fig.~\ref{fig:mzc}). 
However, as we move towards the quantum phase transition, we see a very sharp transverse movement in the 
BVP at $\lambda_{c}=1$. 
From Fig.~\ref{fig:mzc}, it can be seen that the BVP of both $M_z$ and $C_{xx}$ detect 
the quantum phase transition in the model.
We have checked that the BVP of other two-site correlators, including two-site entanglement,
can also detect the QPT in this model.

\subsection{Finite-size scaling}
\label{sec:finsiz}
\begin{figure}[]
\includegraphics[width=0.48\textwidth]{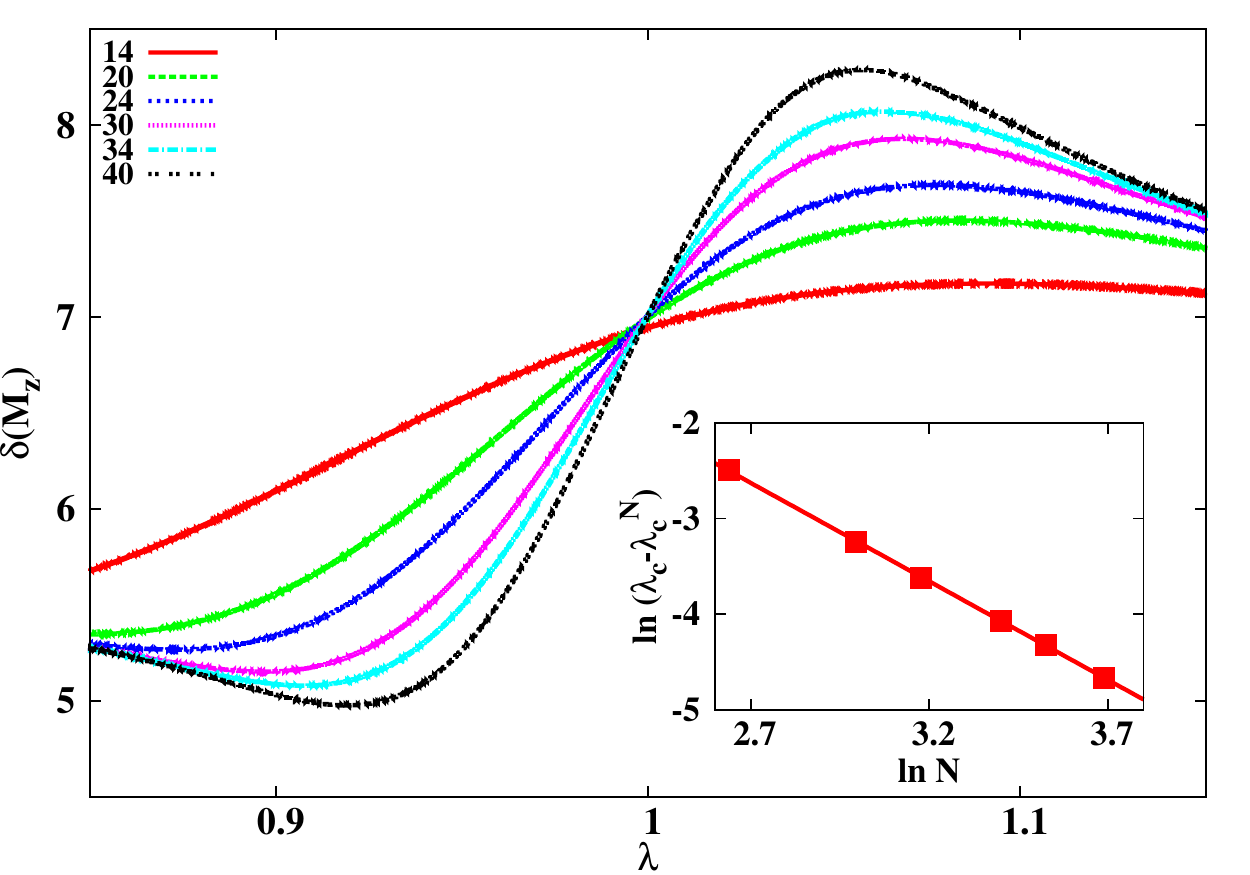}{\centering}
\caption{(Color online.) BVP of transverse magnetization, $\delta(M_z)$, with respect to the 
external magnetic field, with the legends indicating the number of spins.
The violation of Benford's law increases as one moves from the magnetically ordered state for 
$\lambda<\lambda_{c}$, to the paramagnetic state for $\lambda>\lambda_{c}$.
The inset shows the scaling of the critical point ($\lambda_{c}^{N}$) with the system size $N$. 
The data points for the figure in the inset are obtained by using the method illustrated 
in Fig.~\ref{fig:curve_fit}, to get the maximum in the derivative
of $\delta(M_{z})$ with respect to $\lambda$. 
The critical point $\lambda_c^N$ approaches $\lambda_c$ as $N^{-2.06}$.
All axes are dimensionless, except the horizontal one in the inset, which is in 
$\ln$ of the number of spins. The plots and the results are for \(\gamma =0.5\), which remains similar any \(\gamma \ne 0\).
}
\label{fig:bvpmzn}
\end{figure}

By virtue of the current advances in cold gas experimental techniques, one can now engineer finite quantum 
spin systems in laboratories~\cite{Rislam}. It is therefore important to study the QPT point and its
scaling with increasing system sizes for finite systems.
 In our finite-size analysis, we have taken the range of $\lambda$ to be $[0.8,1.2]$ 
 (see Fig. \ref{fig:bvpmzn}). The plot shows the 
BVP as a function of $\lambda$ for different system sizes.
We have considered systems of finite (periodic) chains consisting of $N$ spins, 
with $N=14, 20, 24, 30, 34, 40$. 
The convergence of the BVP is ensured by taking a sufficient number of sample points 
in each subinterval of $\lambda$. Note that the BVP has a large transverse movement around the quantum critical point.
The variation of Benford magnetization with $\lambda$ is similar to the variation of 
transverse magnetization  itself with $\lambda$ around the QPT. 
\begin{figure}[]
\includegraphics[width=0.48\textwidth]{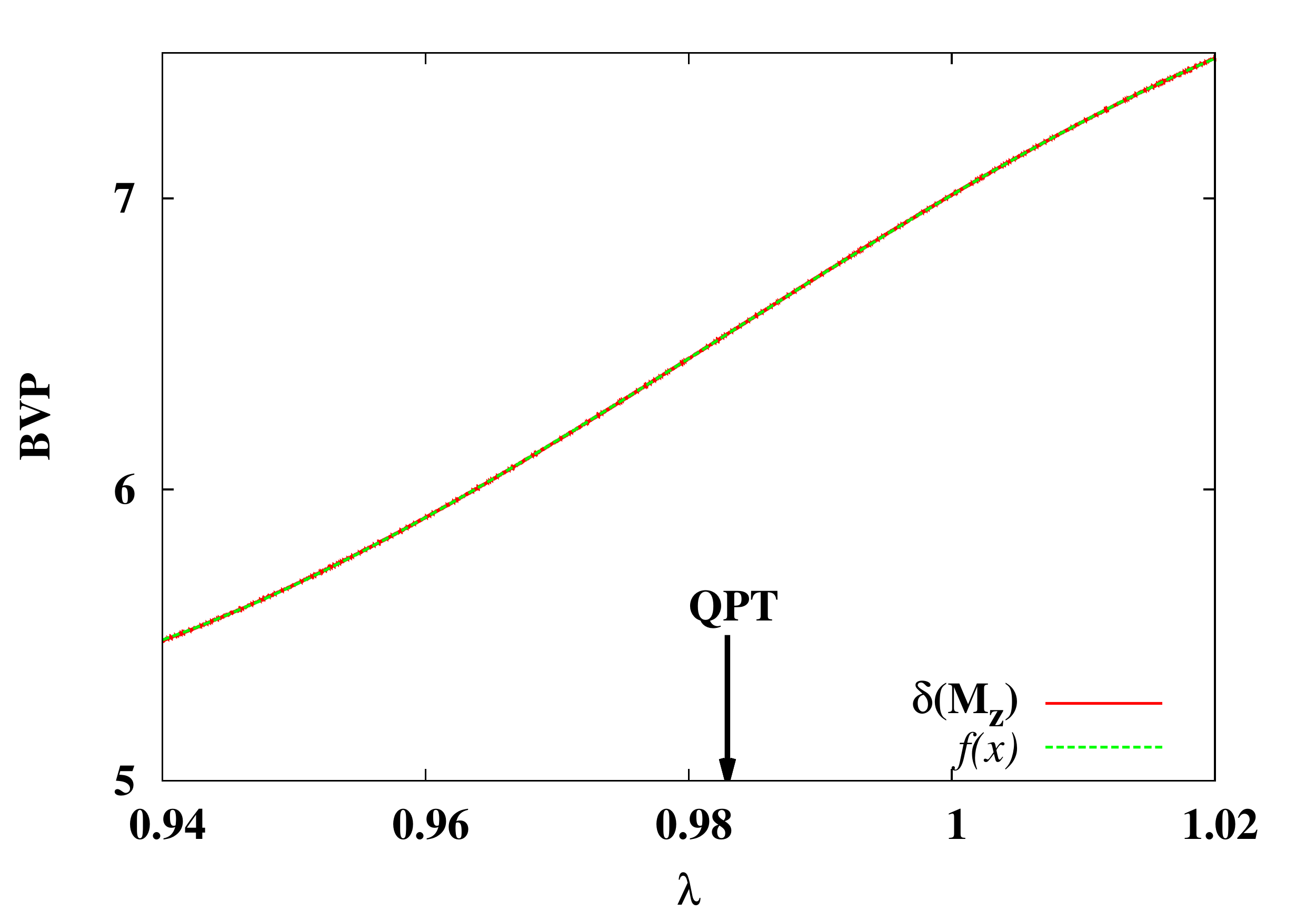}{\centering}
\caption{(Color online.) A cubic polynomial $f(x)$ is fitted to $\delta(M_z)$,
for $N=30$, in the appropriate range of $\lambda$, using
the method of least squares. The constants of $f(x)$ have an error of the order of $0.02$\%.
The point of QPT, marked in the figure at $\lambda=0.9830$, is predicted in this finite system for the considered order parameter,
and corresponds to the maxima of $f{'}(x)$. All axes are dimensionless. 
Note that the two curves, for $\delta(M_z)$ and $f(x)$ have merged with each other. Here, \(\gamma = 0.5\).}
\label{fig:curve_fit}
\end{figure}
It can be readily seen from 
Fig.~\ref{fig:bvpmzn} that the derivative of Benford magnetization with 
$\lambda$ will peak at the point of QPT, as the curvature of the BVP changes from concave to convex there.
Therefore, it is important to find the derivative of BVP with respect to $\lambda$.
The curves in Fig.~\ref{fig:bvpmzn} look quite smooth from a distance, but a 
closer inspection reveals fluctuations in the curves. 
However, it is quite evident that 
the curvatures of the curves change from concave to convex,
around the point of QPT. Four fixed points are required to draw such a curve.
Therefore, we fit a cubic polynomial to the data for BVP in the appropriate range of $\lambda$,
for a fixed $N$, using the 
method of least squares and find the 
exact point where the 
derivative has a maximum. The value of $\lambda$ corresponding to this maxima is
the predicted point of QPT for the particular system size $N$ and the particular order parameter considered.
We denote this value of $\lambda$ by $\lambda_c^N$. We have performed this analysis for 
$N=14, 20, 24, 30, 34, 40$. In Fig.~\ref{fig:curve_fit}, we present a summary of this
analysis for $N=30$.
In the inset of Fig.~\ref{fig:bvpmzn}, we plot $\ln(\lambda_c-\lambda_c^N)$ with respect to $\ln N$.
We find that a straight line fits the plot, which we find via the method of least squares. Exponentiating 
the equation of the straight line, for \(\gamma = 0.5\), we obtain that 
 $\lambda_c^N$ approaches $\lambda_c$ as $N^{-2.06}$ i.e.
\begin{equation}
\lambda^{N}_{c}=\lambda_{c}+kN^{-2.06}.
\end{equation}
\begin{figure}[]
\includegraphics[width=0.48\textwidth]{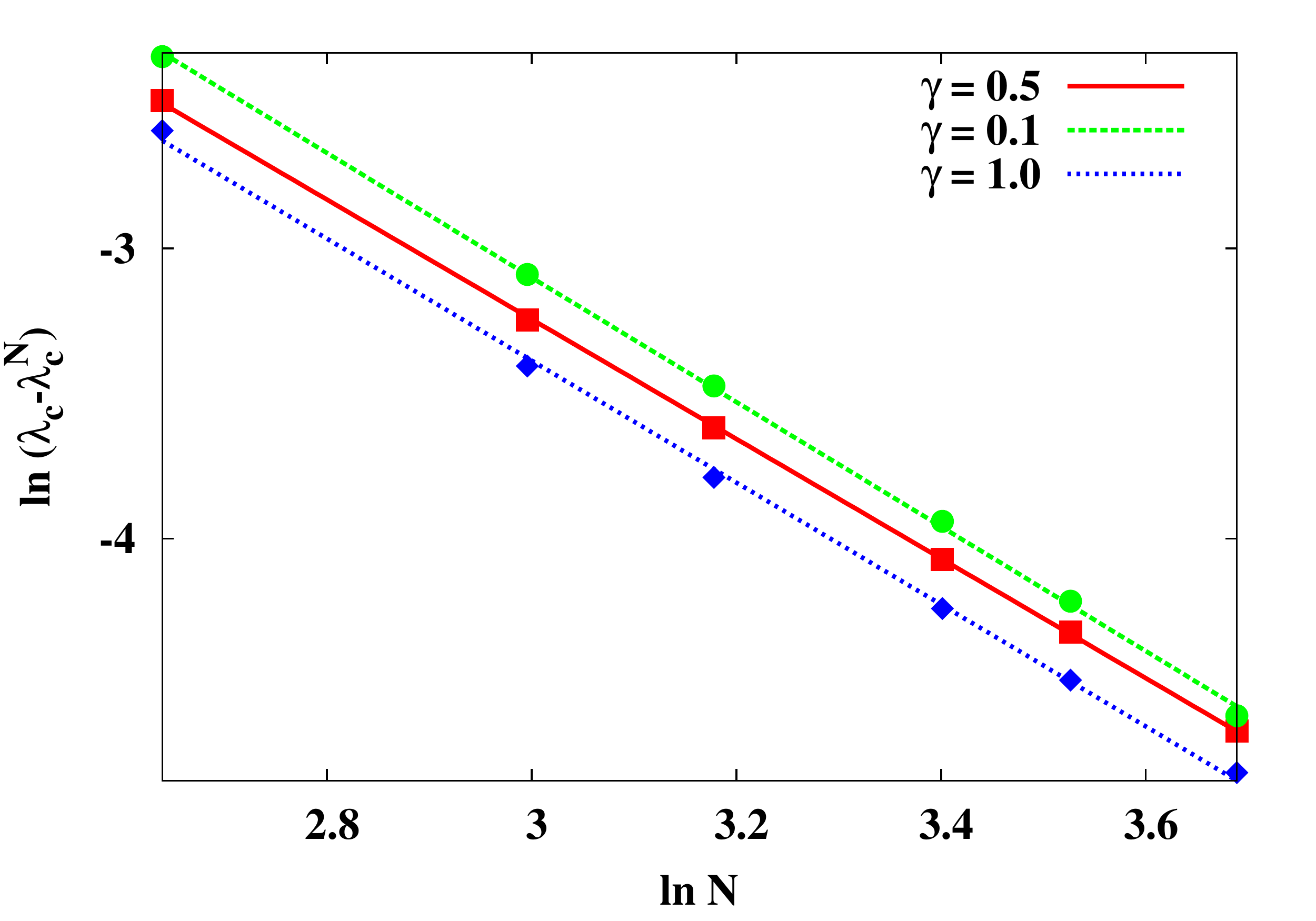}{\centering}
\caption{(Color online.) The scaling of the critical point $\lambda_c^N$ with $N$ for three
values of the anisotropy parameter. The scaling exponents for $\gamma=0.1,0.5$, and $1.0$ are 
-2.14, -2.06, and -2.10 respectively. Note that the lines are almost parallel to each other.
The vertical axis is dimensionless, while the horizontal one is in $\ln$ of the number of spins.}
\label{fig:benf-gamma}
\end{figure} 
The error associated with the estimation of the scaling exponent is of the order of
$0.5$\%.
The scaling exponent found using the Benford magnetization is much higher than many other 
known scaling exponents for this model. In particular,
the scaling exponents for transverse magnetization, fidelity, concurrence, quantum discord, 
and shared purity are 
significantly lower
\cite{fidel_scaling,scaling_ent,shared_pur,classical-Sachdev}. 
In Fig.~\ref{fig:benf-gamma}, we plot the scaling of the critical point $\lambda_c^N$ with $N$
for three values of the anisotropy parameter $\gamma$. The scaling exponents for $\gamma=0.1,0.5$, and $1.0$ are 
-2.14, -2.06, and -2.10 respectively. 

\begin{figure}[h]
\includegraphics[width=0.48\textwidth]{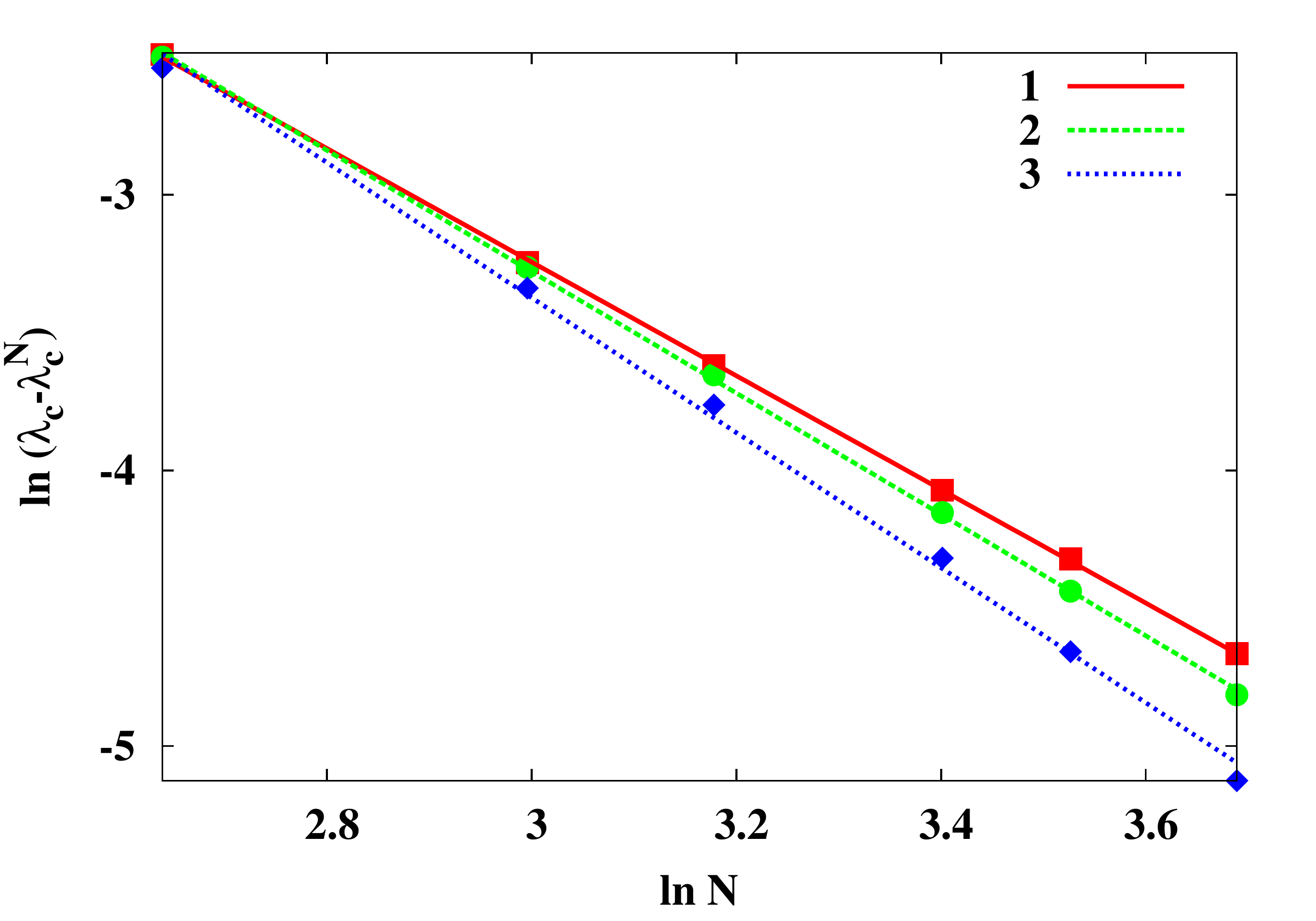}{\centering}
\caption{(Color online.) The scaling of the critical point $\lambda_c^N$ with $N$. The 
Benford distribution has been used in conjunction with the mean deviation (1),
standard deviation (2), and Bhattacharya distance (3), to analyze the data 
for transverse magnetization. The scaling exponents for (1), (2), and (3) are 
-2.06, -2.20, and -2.45 respectively.
The vertical axis is dimensionless, while the horizontal one is in $\ln$ of the number of spins. Here, \(\gamma = 0.5\).}
\label{fig:bf-scale-3m}
\end{figure} 

 Note that the 
 procedure for Benford analysis described in this paper is not unique.
 We compare two frequency distributions,
one of which is obtained from Benford's law and the other is obtained theoretically or 
experimentally from a natural phenomenon.
  The distance
between the two distributions is quantified by a number obtained by using a measure. 
The choice of this measure
is not unique. We have reported the results  until now
by using the mean deviation as our distance measure. But now we also use  
the standard deviation and the Bhattacharya metric~\cite{Bhattacharya} 
to quantify the distance between two frequency 
distributions. So, for any observable $Q$, we define the violation
parameter $\delta(Q)_{sd}$, using the concept of standard deviation, as
\begin{equation}
 \delta(Q)_{sd}= \frac{1}{3}\sqrt{\displaystyle\sum_{D=1}^9 (O_D-E_D)^2}.
\end{equation}
The violation parameter $\delta(Q)_{Bd}$, corresponding to the 
Bhattacharya metric, can be defined as
\begin{equation}
 \delta(Q)_{Bd}=-\ln{\displaystyle\sum_{D=1}^9 \sqrt{O_DE_D}}.
\end{equation}
We observe that the Benford distribution, combined with any of these measures, to
discriminate the observed and expected frequency distributions, 
applied to data obtained for $M_z$ or $C_{xx}$,
can be used to detect the QPT point for the 
infinite quantum XY model. We extend the analysis to finite-sized systems and obtain
the scaling exponents. In Fig.~\ref{fig:bf-scale-3m}, we plot the scaling of the critical 
point $\lambda_c^N$ with $N$, using the Benford distribution in conjunction 
with the mean deviation,
standard deviation, and the Bhattacharya distance, to analyze the data 
for transverse magnetization. We find that the scaling exponents are affected by the 
choice of the measure quantifying 
the violation parameter. The scaling exponents, for \(\gamma = 0.5\), obtained by using the mean deviation, 
standard deviation, and the Bhattacharya distance
are -2.06, -2.20, and -2.45 respectively. This shows that the Bhattacharya metric provides 
a better scaling among the considered metrics for the Benford distribution.

\section{phase transition at Finite temperature}
\label{sec:finite_temp}
In this section, we discuss the finite temperature phase transition
 in the quantum XY model. 
 A quantum phase transition is a phase transition at zero temperature driven by 
quantum fluctuations. But zero temperature is a theoretical concept which
cannot be reached in experiments. However, current cooling methods enable one
to reach temperatures of a few nanoKelvin. Therefore, it is important to study the status of the 
phase transition considered in the preceding section at very low temperatures, 
in the presence of both quantum and thermal fluctuations.
In the finite temperature quantum XY model, the system crosses over from the magnetically
ordered region to the quantum critical region and then to the paramagnetic region
as the field is varied at a fixed finite temperature~\cite{Ssachdev,korpean_book,subir_paper}.

\begin{figure}[]
\includegraphics[width=0.4\textwidth]{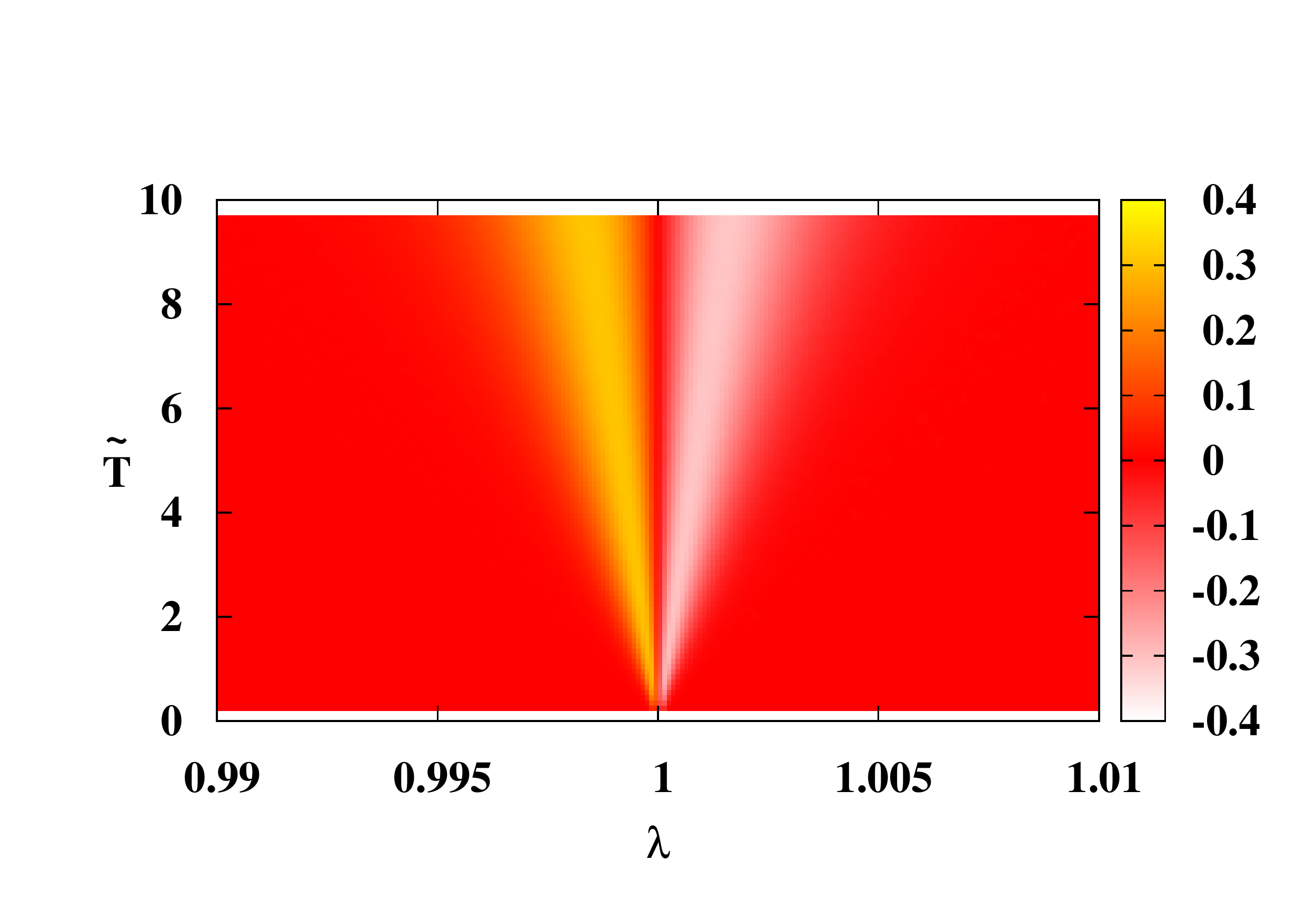}{\centering}
\caption{(Color online.) Projection of the derivative of  
transverse magnetization with respect to temperature, $\partial M_{z}(\lambda,\widetilde{T})/\partial \widetilde{T}$,
on the $\lambda-\widetilde{T}$ plane. The $\widetilde{T}$ axis plotted is multiplied by $10^4$.
The positions of the maxima (yellow) for $\lambda<\lambda_c$ and
minima (white) for $\lambda>\lambda_c$ form the phase transition lines in the $\lambda-\widetilde{T}$ plane.
All quantities plotted are dimensionless. Here, \(\gamma = 1\).}
\label{fig:benford_mag3D}
\end{figure} 
\begin{figure}[]
\includegraphics[width=0.4\textwidth]{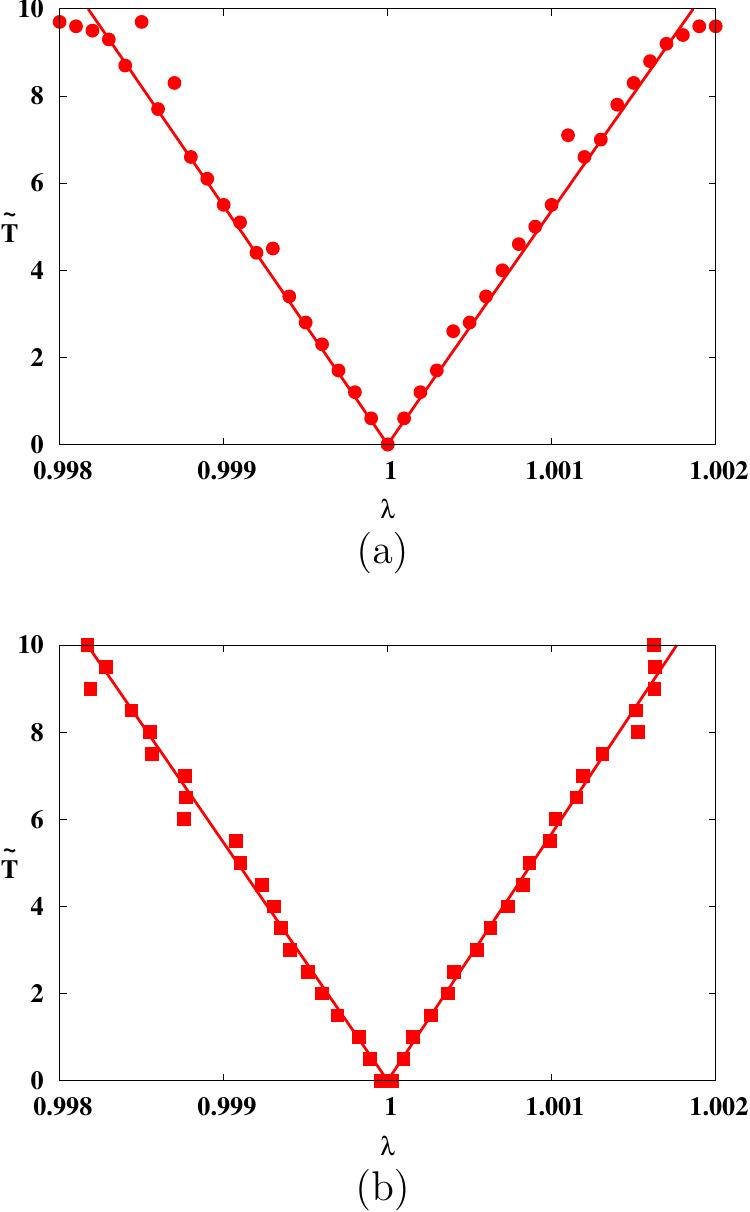}{\centering}
\caption{(Color online.) The cross-over lines are plotted in the ($\lambda-\widetilde{T}$) plane. 
The plots clearly show two phase transitions at finite temperatures at two critical values of the 
driving parameter. The positions of the maxima and minima are plotted for 
$\partial M_{z}(\lambda,\widetilde{T})/\partial \widetilde{T}$ in panel (a), and for 
BVP of transverse magnetization ($\delta(M_z)$) in panel (b). 
All quantities plotted are dimensionless. The $\widetilde{T}$-axes 
plotted are multiplied by $10^4$. The results are for the Ising model (\(\gamma = 1\)), but they remain 
similar for any \(\gamma \ne 0\).}
\label{fig:benford_mag_sys}
\end{figure}
We consider an infinite system at a finite but very low temperature. The transverse magnetization of the 
system is calculated using Eq.~(\ref{eq:fintemp}) as a function of the driving parameter and temperature.
In Fig.~\ref{fig:benford_mag3D}, we plot the projection of the partial derivative of transverse
magnetization with respect to temperature, $\partial M_{z}(\lambda,\widetilde{T})/\partial \widetilde{T}$, in the 
$\lambda-\widetilde{T}$ plane, where $\widetilde{T}=kT/J$, for $\gamma=1$.
The positions of the maxima (yellow) for $\lambda<\lambda_c$ and
minima (white) for $\lambda>\lambda_c$ form the phase transition lines in the $\lambda-\widetilde{T}$ plane.
Note that the temperature is a linear function of the driving parameter at such low temperatures,
along the lines of phase transition. In an effort to investigate the Benford approach further
and compare it with the established methods of data analysis, like the derivative method given above,
we analyze the finite temperature data using Benford's law. We calculate the BVP of transverse magnetization
($\delta(M_z)$),
as a function of the driving parameter for a set of fixed temperatures close to absolute zero.
The data obtained turns out to be very similar to that plotted in Fig.~\ref{fig:benford_mag3D} with 
$\partial M_{z}(\lambda,\widetilde{T})/\partial \widetilde{T}$ replaced by $\delta(M_z)$. We plot the exact 
positions of the maxima and minima on the two panels shown in Fig.~\ref{fig:benford_mag_sys},
which turn out to be almost identical. It can be seen that the system undergoes two phase transitions
at any fixed finite temperature as the driving parameter is varied. Straight lines are fitted to the 
obtained data using the method of least squares. The cross-over lines obtained by using the 
Benford analysis, for \(\gamma = 1\), are given by the equations 
\begin{eqnarray}
  \widetilde{T}&=&-0.546(\lambda-\lambda_{c}),~~~~\lambda<\lambda_{c} \nonumber \\
  \widetilde{T}&=&0.567(\lambda-\lambda_{c}),~~~~~~\lambda>\lambda_{c}. 
\end{eqnarray}
The errors associated with the values of the 
constants in these equations are of the order of 1\%.
We therefore find that just like for the zero temperature case, the finite temperature transitions
are also easily detected by the experimentally less-demanding Benford analysis.

\section{comparison with other distributions}
\label{sec:compare}
\begin{figure}[]
\includegraphics[width=0.4\textwidth]{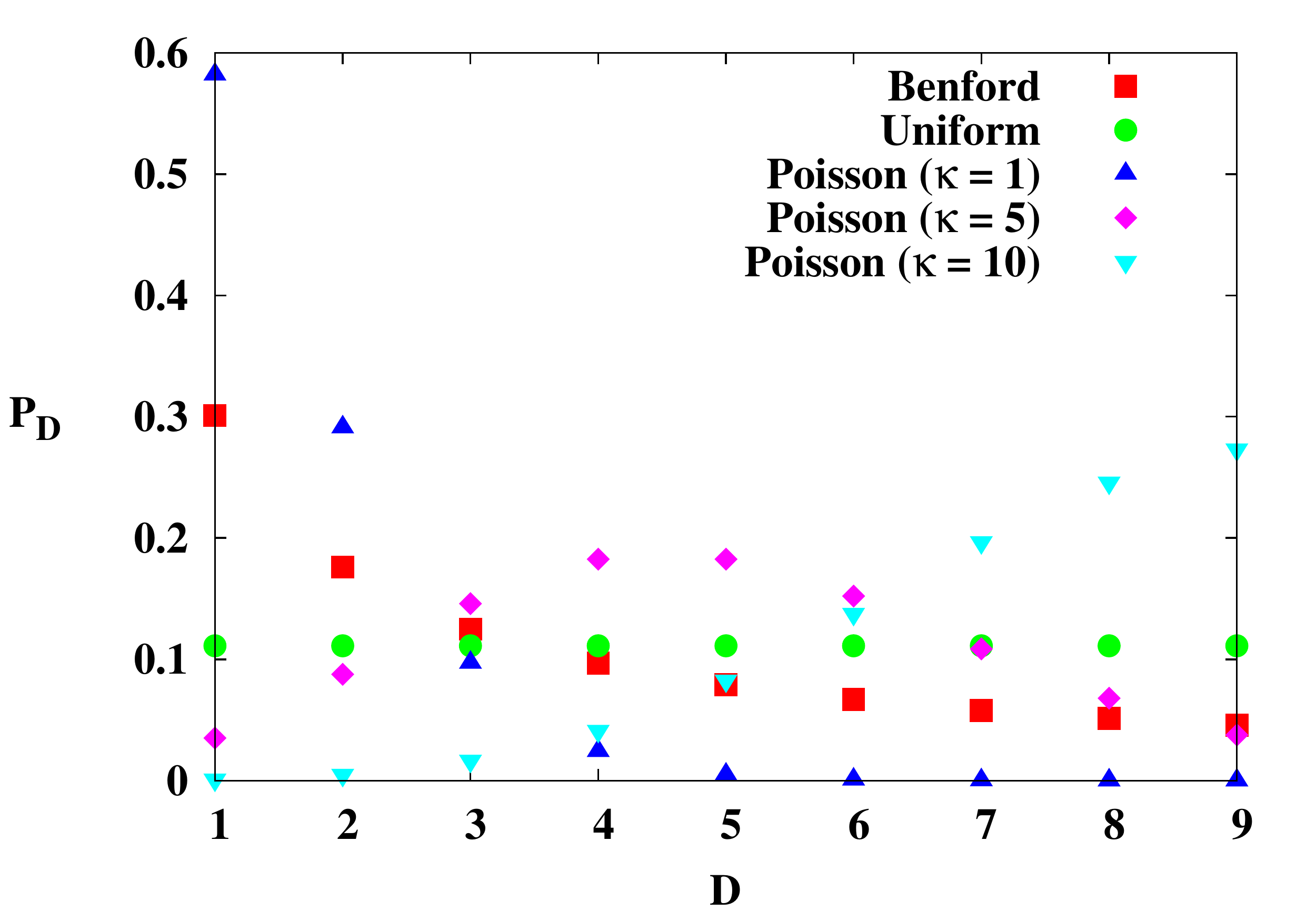}{\centering}
\caption{(Color online.) Frequency of the first significant digits for
the different normalized discrete distributions. Both axes are dimensionless.}
\label{fig:alldis}
\end{figure}

Let us now revert back to the zero-temperature regime. 
Since the BVP detects the QPT point, it is natural to ask if the Benford distribution is the only one that is capable
of detecting the quantum critical point in the quantum XY model. 
Therefore, we investigate the problem using a few other discrete 
frequency distributions. The simplest frequency distribution is the uniform distribution. The frequency
of the first significant digit ``$D$'' of the uniform distribution is given by 
\begin{equation}
 P_D=\dfrac{1}{9}.
\end{equation}
The other discrete distribution that we consider is the Poisson distribution. The  frequency
of the first significant digit ``$D$'' of the Poisson distribution is given by
\begin{equation}
 P_D=\dfrac{\kappa^De^{-\kappa}}{D!}.\dfrac{1}{N_0},
\end{equation}
where the parameter $\kappa>0$. We normalize the distribution by $N_0$, in such a way that
$\sum_{D=1}^9P_D=1$. In Fig.~\ref{fig:alldis}, 
\begin{figure}[]
\includegraphics[width=0.4\textwidth]{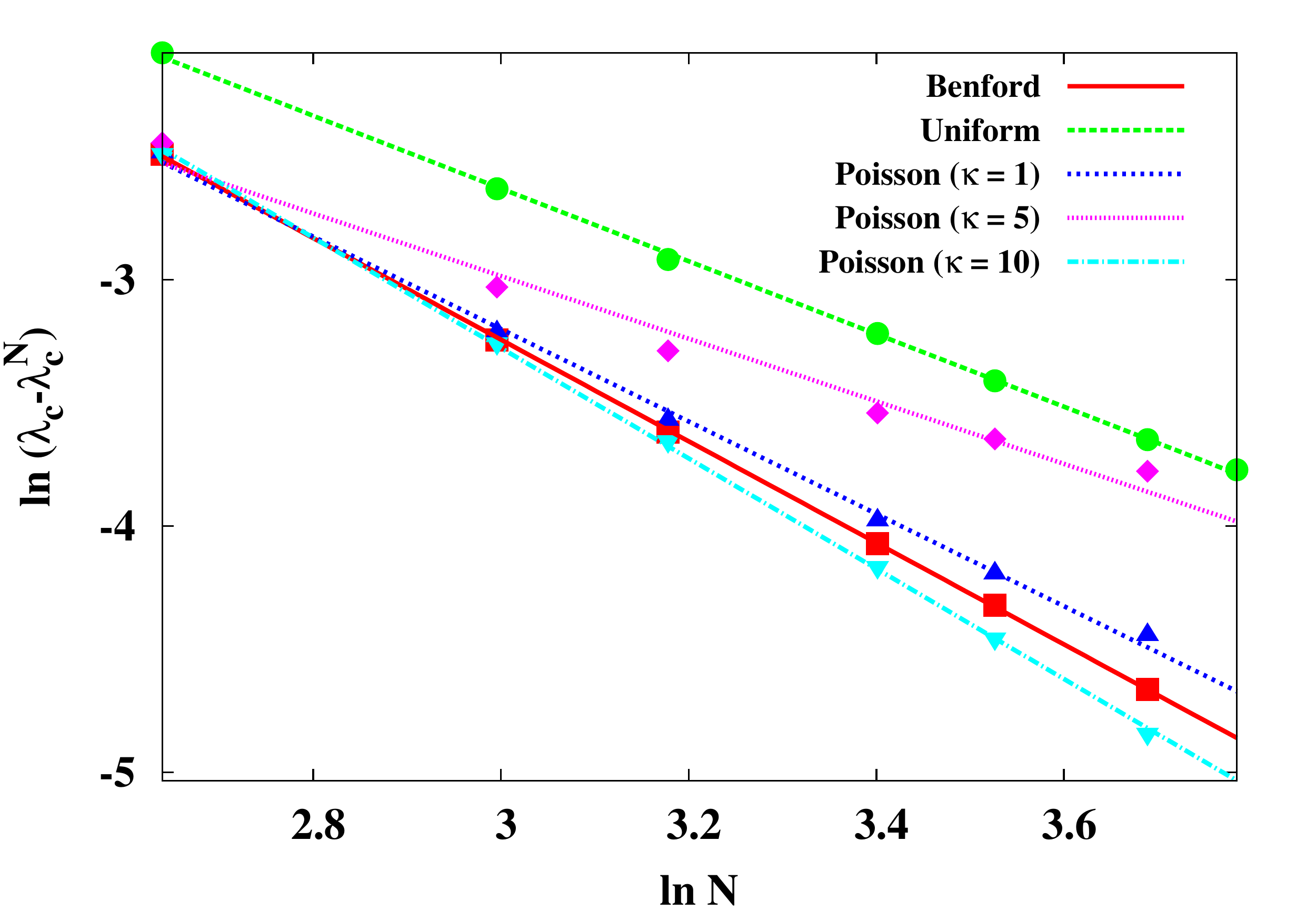}{\centering}
\caption{(Color online.) The scaling of the critical point $\lambda_c^N$ with $N$. The 
indicated frequency distributions have been used in conjunction with the mean deviation
to analyze the data for transverse magnetization. The scaling exponents, for \(\gamma = 0.5\), using
the Benford, uniform, and Poisson ($\kappa=1,5,10$) are 
-2.06, -1.48, -1.88, -1.27, and -2.24 respectively.
The dimensions are as in Fig.~\ref{fig:bf-scale-3m}.}
\label{fig:all-scale-mean}
\end{figure} 
\begin{figure}[]
\includegraphics[width=0.4\textwidth]{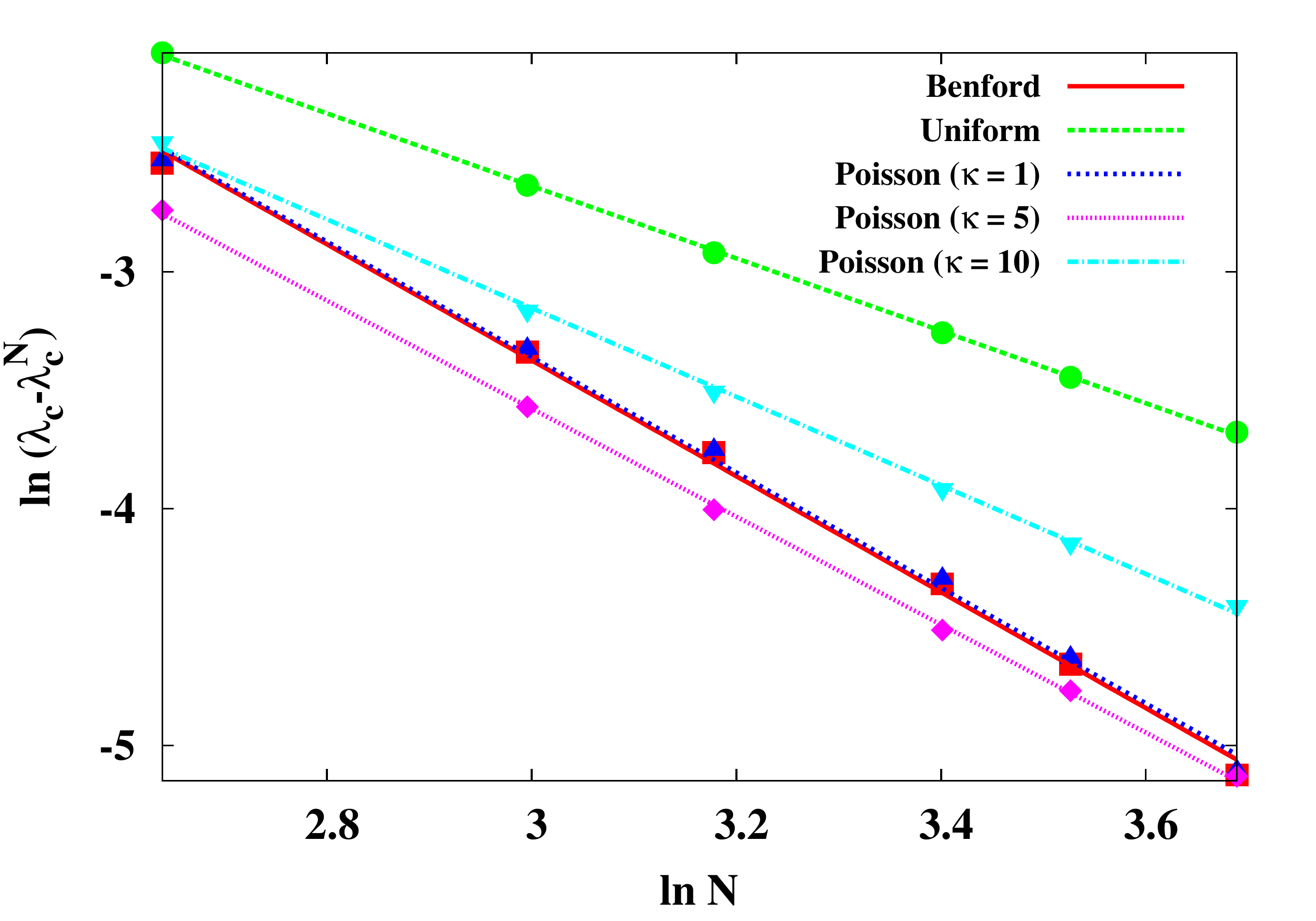}{\centering}
\caption{(Color online.) The scaling of the critical point $\lambda_c^N$ with $N$. The 
indicated frequency distributions have been used in conjunction with the Bhattacharya distance
to analyze the data for transverse magnetization. The scaling exponents, for \(\gamma = 0.5\), using
the Benford, uniform, and Poisson ($\kappa=1,5,10$) are 
-2.45, -1.53, -2.44, -2.28, and -1.87 respectively.
The dimensions are as in Fig.~\ref{fig:bf-scale-3m}.}
\label{fig:all-scale-bd}
\end{figure}
we plot the different frequency distributions. Note that for $\kappa=1$, the Poisson
distribution is qualitatively similar to the Benford distribution, while for $\kappa=10$,
the Poisson distribution is qualitatively similar to the mirror image of the Benford distribution
along the $D=5$ axis. The Poisson distribution is bell-shaped for $\kappa=5$. Now, we 
analyze the theoretical data using the uniform, and the Poisson distributions for $\kappa=1,5,10$. 
We find that any of these distributions in conjunction with any of the three measures, viz. the mean deviation,
the standard deviation, or the Bhattacharya distance, when applied to theoretical data
for $M_z$ or $C_{xx}$ in the infinite quantum XY model, detects the QPT point.
However, the signatures of QPT, obtained by using the 
different distributions
for data analysis, are different. While a minimum of the violation parameter is obtained
for the uniform distribution (at the QPT), a minimum or a maximum in the derivative of the 
violation parameter is obtained for the other distributions.
This signature is independent of the measure used in the analysis.
The data analysis technique is further 
explored by performing  finite-size scaling analysis using the data for transverse magnetization 
for all the different frequency
distributions using the mean deviation and the Bhattacharya 
distance. 
We present the results for \(\gamma =0.5\). However, the results are qualitatively similar for any \(\gamma \ne 0\).
In Fig.~\ref{fig:all-scale-mean}, 
we plot the scaling of the critical point $\lambda_c^N$ with $N$ using all the frequency
distributions discussed, in conjunction with the mean deviation.
We find that the choice of the frequency distribution affects the scaling exponents.
The Poisson distribution with $\kappa=10$ gives the highest scaling exponent of -2.24.
The scaling exponents obtained by using 
the Benford, uniform, and Poisson ($\kappa=1,5$) distributions are -2.06, -1.48, -1.88, 
and -1.27 respectively. 
In Fig.~\ref{fig:all-scale-bd},
 we plot the scaling of the critical point $\lambda_c^N$ with $N$, again using all the frequency
distributions, but now, in conjunction with the Bhattacharya distance. 
The scaling exponents obtained using the Benford, uniform, and Poisson ($\kappa=1,5,10$) 
distributions are -2.45, -1.53, -2.44, -2.28, and -1.87
 respectively. Note that the scaling exponents obtained using the Benford distribution
is among the higher ones for any measure and for any distribution. Therefore, while the 
procedure for data analysis involving the Benford's law is not unique, there seems to be some
evidence that it is one of the better ones, if not the best that can be employed to analyze the given 
data.

\section{Conclusion}
\label{sec:conclusion}
The Benford's law seems to be a very efficient tool in analyzing data. Our data analysis 
and subsequent results, for a
quantum mechanical system reinforces the belief. We find that the quantum phase transition in the one-dimensional
anisotropic quantum XY models, are efficiently detected by this analysis.
The main advantage of this technique is its dependence on only the first significant digit of an observable.
This advantage is very important from the point of view of experiments, in which accuracy is not very high.

 We have used a number of other discrete frequency distributions along
with three different measures to discriminate between frequency distributions. It is seen that the Benford 
distribution is not unique and that other distributions can also be employed to analyze data and detect 
a quantum critical point in the quantum XY models. However, analysis of data using the Benford distribution
produces higher scaling exponents in almost all cases. This indicates that this is possibly
a better tool for data analysis. In this paper, we have compared pairs of frequency distributions,
one of which is obtained from quantum theory and the other is obtained from a probability distribution. 
The distance
between the two distributions is quantified by a number obtained by using a measure. The choice of this measure
is not unique. We have reported our results by using the mean deviation, 
standard deviation, and the Bhattacharya metric to quantify the distance between the two frequency 
distributions.
The analysis has been carried out for both finite 
and infinite size systems for different observables at zero and finite temperatures. 
Interestingly, most of the finite size scaling exponents obtained in this paper
by using the violation parameters 
are much higher than those obtained using other measures 
like magnetization, fidelity, concurrence, shared purity, and quantum discord.
We also find the linear relationship between temperature and 
the driving parameter at a finite temperature, along the cross-over lines, in this model.
The analysis strongly suggests that measuring the Benford violation parameter in the 
laboratory can be an efficient tool for detecting phase transitions in quantum many-body systems.

\end{document}